\documentclass[showpacs,preprintnumbers,floatfix,12pt,twocolumns]{revtex4}
\usepackage{graphicx}
\usepackage{amsmath}
\usepackage{float}
\usepackage{caption}
\usepackage{rotating}

\begin{document}

\title{Probability flux as a method for detecting scaling}
\author{M. Ignaccolo}
\affiliation{Physics Department, Duke University, Durham (NC)}
\author{P. Grigolini}
\affiliation{Center for Nonlinear Science, University of North Texas, Denton (TX)}
\author{B. J. West}
\affiliation{Information Science Directorate, Army Research Office, Durham (NC)}
\date{\today }

\begin{abstract}
We introduce a new method for detecting scaling in time series. The method
uses the properties of the probability flux for stochastic self-affine
processes and is called the \textit{probability flux analysis} (PFA). The
advantages of this method are: 1) it is independent of the finiteness of the
moments of the self-affine process; 2) it does not require a binning
procedure for numerical evaluation of the the probability density function.
These properties make the method particularly efficient for heavy tailed
distributions in which the variance is not finite, for example, in L\'{e}vy $%
\alpha -$stable processes. This utility is established using a comparison
with the \textit{diffusion entropy} (DE) method.
\end{abstract}

\maketitle

\section{\protect\bigskip Introduction}

\label{intro} Over the past ten years there has been an explosion of
research papers published in the area of complex networks. Some would argue
that this torrent of publications is a continuance of the growing awareness
of complexity science whose origin can be traced back to the decade of the
1960s. The present emphasis on scale-free networks and their implications
for scientific disciplines from sociology to neurophysiology had its
beginnings with the fractal time series analysis of Mandelbrot \cite%
{mandelbrot} and the scaling parameter method of Hurst \cite{hurst}. From
this early appreciation for the limitations of 'normal' statistics to
explain and/or characterize complex phenomena we have such terms as
'self-similar', 'self-affine', 'scale invariant', 'fractal', and
'multi-fractal' to capture the non-homogeneous and non-isotropic behavior of
statistical variability \cite{feder,schroeder}. As the mathematical
developments became more familiar to a generation of scientists the
investigations into the manifestations of these effects in biology,
economics, geophysics, hydrology, neurophysiology, sociology and so on
steadily increased in number. First there were a few isolated studies,
followed by a steady rate of overlapping investigations, resulting in what
is now a tsunami of monographs, technical papers and popular articles.
Consequently, even though no universally accepted definition of complexity
has emerged, the consensus of scientific opinion has converged on the use of
scaling as one signature of complexity.

The scientific investigations into scaling have historically been of two
kinds: 1) the formal mathematics identifying the properties data sequences
must possess in order to scale in well-defined ways along with the
techniques to analyze the data and reveal that scaling and 2) the
application of those techniques to time series measured in complex
phenomena. A stochastic process $X(t)$ is said to scale if the
time-dependence of the random variable is such that $X(t)$$=$$K^{-\delta }$$%
X(Kt)$ with $K>0$ and $\delta $ is a scaling parameter. Statistical methods
to detect self-affinity (scaling) in time series identify the variable $X(t)$
with the integrated time series and not with the time series itself $\xi
\left( t\right) $, such that,

\begin{equation}
X(t)=\overset{t}{\underset{0}{\int }}\xi \left( t^{\prime }\right)
dt^{\prime }.  \label{diffusion1}
\end{equation}%
Thus, the original time series $\xi \left( t^{\prime }\right) $ may be
viewed as a sequence of increments of the variable $X$. In this sense many
of the data analysis methods are \textquotedblleft
diffusive\textquotedblright\ in that the variable $X$ is the aggregation of
fluctuations denoted by the integral (\ref{diffusion1}). Since its
introduction in the middle 1990s \textit{detrended fluctuation analysis}
(DFA) \cite{peng} has become the method of choice for detecting scaling
particularly in biomedical time series \cite{bassingthwaighte}. The
motivation for introducing DFA was the presumed applicability of the method
to non-stationary time series, particularly those with long-time
correlations. The scaling methods discussed by Mandelbrot \cite{mandelbrot}
and DFA are \textquotedblleft variance\textquotedblright\ methods in that
they assume the time dependence of the variance of the stochastic variable $%
Var(X(t))$ with scaling parameter $\delta $ is algebraic, namely:

\begin{equation}
Var(X(t))\propto t^{2\delta }.  \label{diffusion2}
\end{equation}

However there are scaling processes such as L\'{e}vy flights \cite%
{shlesinger} for which such relationships are violated because the second
moment diverges or L\'{e}vy walks \cite{shlesinger} for which the second
moment is finite and does satisfy a scaling relation similar to (\ref%
{diffusion2}). The necessity for detecting the proper scaling of L\'{e}vy
processes was one reason behind the development of the \textit{diffusion
entropy analysis} (DEA) method \cite{dea} described in Section II. 
The evident advantage of the information or diffusion entropy $S(t)$ over
second moment methods is that the former is always finite, independently of
the behavior of the moments of the probability density distribution (\textit{%
pdf}) $p(x,t)$. The quantity $p(x,t)dx$ is the probability of finding the
trajectory $X(t)$ in a infinitesimal neighborhood of $x$ at time $t$. The
divergence of the central moments, typical of L\'{e}vy processes, create
difficulties in the numerical determination of the $pdf$ and the associated
entropy.

Herein we propose a new procedure for determining scaling, the \textit{%
probability flux analysis} (PFA) as a general method for scaling detection.
Since the PFA uses the cumulative probability instead of the density $p(x,t)$
itself as such it is statistically more efficient than DEA. The cumulative
distribution integrates the \textit{pdf} and reduces the noise due to the
adoption of a statistical ensemble with a finite number of trajectories. The
PFA method is shown to outperform the other methods in the case of L\'{e}vy
processes. The present paper is structured as follows. In Section~\ref%
{defofscaling} we introduce scaling for the \textit{pdf} $,$ which is an
extension of the concept of self-affinity. Ordinarily self-affinity is
recovered in the algebraic time dependence of a scaling function $\beta (t)$
as we show using the DEA method and its numerical implementation. We
introduce the PFA method in Section~\ref{pfamethod}, and compare its
performance with that of DEA on computer generated time series in Section~%
\ref{pfavsde} and subsequently make the same comparisons with real world
data. Finally, we draw some conclusions in Section~\ref{conclusion}.

\section{Scaling}

\label{defofscaling} \label{scacondi} Consider a one-dimensional stochastic
trajectory $X(t)$ whose statistical properties are described by $p(x,t)$.
The stochastic process represented by $X(t)$ is said to scale if the
corresponding \textit{pdf} satisfies the scaling relation 
\begin{equation}
p(x,t)=\frac{1}{\beta \left( t\right) }F\left( \frac{x}{\beta \left(
t\right) }\right)  \label{scaling}
\end{equation}%
where the scaling function $\beta (t)$ is a function of time and without
loss of generality we have assumed $X(0)=0$. The function $F$ in the above
equation is the scaled density since 
\begin{equation}  \label{scaleddensity}
\int p(x,t)dx=\int F(y)dy=1.
\end{equation}
The relation~(\ref{scaling}) proposed here is a generalization of the
widespread notion of scaling adopted in the literature \cite{feder,schroeder}%
, which often corresponds to the particular case 
\begin{equation}
\beta (t)=t^{\delta }.  \label{algebraicscaling}
\end{equation}%
Herein we refer to the validity of the scaling relation~(\ref{scaling})
together with (\ref{algebraicscaling}) as the \textquotedblleft
algebraic\textquotedblright\ scaling condition. Geometrically, the scaling
condition implies that $\ p(x,t)$ is invariant under the transformations 
\begin{equation}
\left\{ 
\begin{array}{l}
x\rightarrow \frac{\beta (t)}{\beta (Kt)}x \\ 
t\rightarrow Kt%
\end{array}%
\right. \;\;\;\;\;\Leftrightarrow \;\;\;\;\;X(t)\underset{s}{=}\frac{\beta
(t)}{\beta (Kt)}X(Kt)  \label{selfaffine}
\end{equation}%
where $K$$>$$0$ and the symbol $\underset{s}{=}$ denotes equality in the
sense that the \textit{pdf}s for the variables on either side of the equal
sign are the same. In the case of algebraic scaling ($\beta (t)$$=$$%
t^{\delta }$) we have 
\begin{equation}
\left\{ 
\begin{array}{l}
x\rightarrow K^{-\delta }x \\ 
t\rightarrow Kt%
\end{array}%
\right. \;\;\;\;\;\Leftrightarrow \;\;\;\;\;X(t)\underset{s}{=}K^{-\delta
}X(Kt).  \label{selfaffinealg}
\end{equation}%
This last set of relationships defines a self-affine transformation \cite%
{feder}.

\subsection{Scaling detection with Diffusion Entropy Analysis (DEA)}

\label{demethod} \label{diffusionentropy} The \textit{pdf} defined for the
stochastic process $X(t)$ can be used to calculate the information entropy.
This use of entropy was implemented in discrete form for coding information
by Shannon \cite{shannon} and is referred to as the Shannon entropy.
Cotemporaniously, this use of entropy was introduced in continuous form by
Wiener for studying the problem of filtering noise from messages in
electrical circuits \cite{wiener}. In the analysis here we use the
continuous form of information entropy 
\begin{equation}
S(t)=-\int p(x,t)\log _{2}p(x,t)dx,  \label{entropy}
\end{equation}%
which was originally identified as diffusion entropy by Scafetta \textit{et
al}. \cite{dea} as a tool for detecting scaling in time series. The
advantages of using entropy rather than the variance to detect scaling is
that entropy provides a more complete description of the stochastic process.
The two approaches become equivalent only when the \textit{pdf} is Gaussian.
In the general case the distribution is not Gaussian and the central moments
can and do diverge, as in the case of\textit{\ }$\ \alpha -$stable L\'{e}vy
distributions . If the scaling condition (\ref{scaling}) is satisfied then
it is straightforward to show that the entropy reduces to 
\begin{equation}
S(t)=S_{0}+\log _{2}\beta (t),  \label{entropyscaling}
\end{equation}%
while in the case of algebraic scaling, given by Eq.~(\ref{algebraicscaling}%
), we have 
\begin{equation}
S(t)=S_{0}+\delta \log _{2}t,  \label{entropyscalingalg}
\end{equation}%
where the additive constant is defined by the integral over the scaled
variable

\begin{equation}
S_{0}=-\int F(y)\log _{2}F(y)dy.  \label{constant}
\end{equation}

The empirical determination of the histograms\ replacing the \textit{pdf}'s
and the discretization of the integral~(\ref{entropy}) is done as follows. A
discrete realization $X_{l}$ ($l$$=$0,1,2,...,$N$, and $X_{0}$$=$$0$) of the
stochastic process $X(t)$ is used to create the set of trajectories 
\begin{equation}
\left\{ X_{k}(t)\right\} =\left\{ X_{k+t}-X_{k}\right\} \;\;k=0,1,2,...,N-t.
\label{pseudogibbs}
\end{equation}%
We call the set $\left\{ X_{k}(t)\right\} $ of $N-t+1$ trajectories the
Single Trajectory Ensemble (STE) as distinct from the Multiple Trajectory
Ensemble (MTE) generated using $N-t+1$ different realization of the
stochastic process $X(t)$. The rationale for the STE is that in many real
world applications one has only a single realization of the stochastic
process available. The two ensembles are generally thought to produce
identical results when the stochastic process is stationary and ergodic;
even in the non-stationary case the STE and MTE averages are thought to be
the same provided the effect of local trends can be eliminated, \textit{e.g}%
., by using the DFA algorithm~\cite{peng}. However, the equivalence between
STE and MTE is lost even in the stationary and ergodic case \cite{newus1},
and caution must be used when interpreting the results of scaling analysis
that rely on the STE. The procedure described in Eq.~(\ref{pseudogibbs}) to
generate the STE has been called the \textquotedblleft
overlapping\textquotedblright\ windows method as two trajectory $X_{k_{1}}(t)
$ and $X_{k_{2}}(t)$ may share a common profile if $\mid $$k_{2}$$-$$k_{1}$$%
\mid $$<$$t$. The overlapping windows method is often preferred to the
non-overlapping windows methods because of the larger number of trajectories
produced \cite{dereport}. 

In the present manuscript the STE is used to calculate the histogram for the 
\textit{pdf} $p_{j,\Delta }(t)$ for finding a trajectory within an interval
of size $\Delta $ centered on the value $x_{j}$. The bin size $\Delta $ has
to be sufficiently small to consider the \textit{pdf} constant within the
interval $[x_{j}-\Delta /2,x_{j}+\Delta /2]$, and the integral of Eq.~(\ref%
{entropy}) can then be approximated by the sum 
\begin{equation}
S(t)\simeq -\sum\limits_{j=1}^{{}}p_{j,\Delta }(t)\log_{2} [p_{j,\Delta
}(t)]+\log_{2} \Delta .  \label{entropynumeric}
\end{equation}
The accuracy of this numerical approximation decreases as $t$ increases
since we have only $N-t+1$ of trajectories in the STE. Values of the density 
$p(x,t)$$\sim$1/$(N-t+1)$ are impossible to reproduce correctly. Moreover,
if the stochastic process $X(t)$ is such that the probability of observing
large values, positive and/or negative, increases in time than the \textit{%
pdf} will assumes increasingly smaller values furhter compromising the
validity of Eq.~(\ref{entropynumeric}). This effect is particularly
drammatic when the stochastic process has infinite second and and/or first
moment such as the L\'{e}vy flights and walks considered in Section~\ref%
{pfavsde}.


\section{Scaling detection with Probability Flux Analysis (PFA)}

\label{pfamethod} The rationale for the \textit{probability flux analysis}
(PFA) is to have a method of scaling detection that is independent of the
binning procedure used to evaluate the histogram for the \textit{pdf},
statistically more accurate, and independent of the size of the moments of
the distribution. Define a constant in the interval $\theta \in $$]0,1[$ and 
$x_{\theta }(t)$ to be a number such that 
\begin{equation}
\theta =\int\limits_{-\infty }^{x_{\theta }(t)}p(x^{\prime },t)dx^{\prime
}\equiv \mathcal{P}(x_{\theta}(t),t)\;\;\forall t,  \label{pfadefinition1}
\end{equation}
where $\mathcal{P}(x,t)$ is the cumulative distribution. The value of the
variate $x_{\theta }(t)$ encompasses a fraction $\theta$ of the probability
density $p(x,t)$. We call PFA any algorithm that at any time step $t$
calculates the number $x_{\theta }(t)$.

If the scaling condition on the \textit{pdf} (\ref{scaling}) is satisfied, 
\begin{equation}
\mathcal{P}(x,t)=\int\limits_{-\infty }^{x}p(x^{\prime },t)dx^{\prime
}=\int\limits_{-\infty }^{x}\frac{1}{\beta (t)}F\left( \frac{x^{\prime }}{%
\beta (t)}\right) dx^{\prime }=\mathcal{F}\left( \frac{x}{\beta (t)}\right) 
\label{pfascaling1}
\end{equation}%
where the function $\mathcal{F}$ is the cumulative distribution of the
scaled density $F$. Thus, for a scaling \textit{pdf} the condition (\ref%
{pfadefinition1}) becomes 
\begin{equation}
\theta =\mathcal{F}\left( \frac{x_{\theta }(t)}{\beta (t)}\right)
\;\;\forall t\;\Rightarrow \;x_{\theta }(t)=z_{\theta }\beta (t)
\label{pfascaling2}
\end{equation}%
where $z_{\theta }$ is a constant. This equation shows that in the case of
scaling the motion in time of the location $x_{\theta }$ with a fraction $%
\theta $ of the trajectories $X(t)$ on its left side is directly
proportional to the motion described by the scaling function $\beta (t)$. In
case of algebraic scaling, Eq.~(\ref{scaling}) and Eq.~(\ref%
{algebraicscaling}), 
\begin{equation}
x_{\theta }(t)=z_{\theta }t^{\delta }\;\Rightarrow \;\ln [x_{\theta
}(t)]=\ln (z_{\theta })+\delta \ln (t)  \label{pfaalgebraicscaling}
\end{equation}%
In the case of a constant drift, Eqs.~(\ref{pfadefinition1})$-$(\ref%
{pfaalgebraicscaling}) are valid in the moving reference frame centered on $%
\omega t$, or equivalently for the variable $x_{\theta }(t)$$-$$\omega $$t$
instead of $x_{\theta }(t)$. As an example, let us consider the stocastic
process $X(t)$ to be Browninan motion with no drift, then the \textit{pdf} $%
p(x,t)$ is a Gaussian function centered on the origin $x$$=$$0$, while the
scaling function $\beta (t)$$=$$\sqrt{t}$. Consider $\theta $$=$0.977 that
is $x_{0.977}$ encompasses 97.7\% of the distribution. For a Gaussian
distribution $x_{0.997}$ correspond to a value in excess of two standard
deviation from the mean. Thus for a Brownian motion with no drift $%
x_{0.997}(t)$$=$2$\sigma $$\sqrt{t}$ which satisfies the second relation of
Eq.~(\ref{pfascaling2}) with $z_{\theta }$$=$2$\sigma $, $\sigma $ being the
standard deviation of the increments of $X(t)$.

The numerical calculation of the function $x_{\theta }(t)$ can be done as
follows. At any time $t$ the trajectories $\left\{ X_{k}(t)\right\} $ of the
STE defined in Eq.~(\ref{pseudogibbs}) are placed in ascending order and the
value $x_{\theta }(t)$ is assigned to be $X_{K}(t)$ with $K$=$[\theta \times
(N-t+1)]$, $[..]$ indicating the integer part. This procedure requires no
binning as the DEA algorhitm, however it can be computationally demanding
since at any time step $t$ the trajectories $\left\{ X_{k}(t)\right\} $ must
be sorted. Hereby, we use a faster procedure. We estimate $x_{\theta }(t)$
by calculating the cumulative distribution $\mathcal{P}(x,t)$ at fixed
spatial intervals of length $\Delta $: the accuracy (resolution) with which
the value $x_{\theta }(t)$ is computed. Let $\left\{ X_{k}(t)\right\} $ be
the set of trajectories of the STE at time $t$, and $M(t)$ the total number
of bins of length $\Delta $ necessary to cover the span of these
trajectories, we define $x_{\theta }(t)$ as follows 
\begin{equation}
x_{\theta }(t)=x_{l}\;\;\text{such that}\;\;\sum\limits_{j=1}^{l}p_{j,\Delta
}(t)<\theta \;\;\text{and}\;\;\sum\limits_{j=l+1}^{M(t)}p_{j,\Delta
}(t)>\theta   \label{xalphanumeric}
\end{equation}%
where the symbol $p_{j,\Delta }(t)$ indicates the trajectory frequency or
histogram within the $j$-th interval $[x_{j}-\Delta /2,x_{j}+\Delta /2]$.
Calculating the frequencies $p_{j,\Delta }(t)$ requires a single sequential
scanning of the trajectories $\left\{ X_{k}(t)\right\} $ while any sorting
algorithm requires more computational power. The procedure described by Eq.~(%
\ref{xalphanumeric}) is similar to the DEA algorithm (\ref{entropynumeric})
as both methods require the calculation of $p_{j,\Delta }(t)$. However,
there is a fundamental difference. For the DEA to be meaningful $p_{j,\Delta
}(t)$/$\Delta $ must accurately reproduce the \textit{pdf} in each interval.
This strong requirement is not necessary for the method described in (\ref%
{xalphanumeric}). The adoption of a binning procedure in the case of PFA is
just a computational device to speed up calculation.

\section{Comparison between PFA and DE method}

\label{pfavsde} In this section, we compare the results of the PFA and DEA
methods applied to a number of computer generated sequences having known
statistical properties. In addition we apply the two techniques to the
electroencephalogram (EEG) data set previously analyzed \cite{massiEEG}.

\subsection{L\'{e}vy Flights}

In this section we generate a number of L\'{e}vy flights for the stochastic
process $X(t).$ These processes have stationary delta-correlated increments $%
\xi (t)$ with infinite variance and possibly also infinite mean depending on
the value of the L\'{e}vy index selected. The generalized central limits
theorem predicts the \textit{pdf} to be, after an initial transient that
depends on the specific distribution of the increments, a L\'{e}vy
distribution. The function $F(y)$ on the right hand side of Eq.~(\ref%
{scaling}) is a stable L\'{e}vy distribution whose Fourier transform is
given by 
\begin{equation}
\mathcal{F}(k)=\exp \left[ ~ik\gamma \!-\!|ck|^{\alpha }\,(1\!-\!i\eta \,%
\text{sgn}(k)\Phi )~\right] .  \label{levy}
\end{equation}%
In the above equation $\mathcal{F}(k)$ is the characteristic function of $F(y)$, $\gamma \in \mathcal{R}$ 
is the shift parameter, $\eta \in $[-1,1] is
called the skewness parameter, a measure of asymmetry, and $0<\alpha \leq 2$
is the L\'{e}vy parameter. Finally, the constant $\Phi $ is equal to $\tan
(\pi \alpha /2)$ for all values of $\alpha $ except for $\alpha $$=$1 when $%
\Phi $$=$$(2/\pi )\log |k|$.

We choose the increments $\xi $ of the variable $X$ to be distributed
according to an inverse power law 
\begin{equation}
\psi ({\xi })=\frac{(\mu -1)B^{(\mu -1)}}{(B+\left\vert \xi \right\vert
)^{\mu }}  \label{plaw}
\end{equation}%
where $\mu $$\in $]1,$+\infty $[ is the \textquotedblleft
index\textquotedblright\ of the inverse power law and $B\in \mathcal{R}$ is a
location parameter ($\xi \gg B\Leftrightarrow \psi (\xi )\propto \left\vert
\xi \right\vert ^{-\mu }$). The scaling parameter $\delta$ of the resulting L\'{e}vy distribution for $p(x,t)$, 
the power law index $\mu$, and the L\'{e}vy parameter $\alpha$ are connected 
as follow: 
\begin{equation}
\alpha =\mu -1 \;\;\; \textrm{and} \;\;\; \delta=\frac{1}{\alpha}=\frac{1}{\mu-1} \;\;\;\;
\textrm{for }\mu \in ]1,3[.  \label{alphamudelta}
\end{equation}
where, of course, $\alpha$$=$2 corresponds to the Gaussian distribution,
which we do not consider here. Random numbers distributed according to (\ref%
{plaw}) can be obtained from random numbers uniformly distributed in the
interval ]0,1[ \cite{palabuiatti}. To test the performance of PFA and DEA on
L\'{e}vy flights, we generate $10^{7}$ random numbers distributed according
to the inverse power law Eq.~(\ref{plaw}) with location parameter $B=1$ and 
power law index $\mu =1.6$. The result are recorded in figures below.

We see from the calculation depicted in Figure 1 that both the DEA and PFA
methods apparently provide reliable estimates of the early time scaling of
the stochastic process, even though we have only used the first 25\% of the
data in the use of PFA. The slope of the curves in both calculation is $1.%
\overline{6}$, the expected scaling index $\delta$ in (\ref{alphamudelta}). 
However after two decades the DEA curve
begins to run out of statistics, whereas the PFA persists for another decade
and one-half before it markedly deviates from the theoretical curve. The
detailed divergence between the two calculations is evident in the insert.

\begin{figure}[ht]
\includegraphics[angle=-90,width=1.0\linewidth]{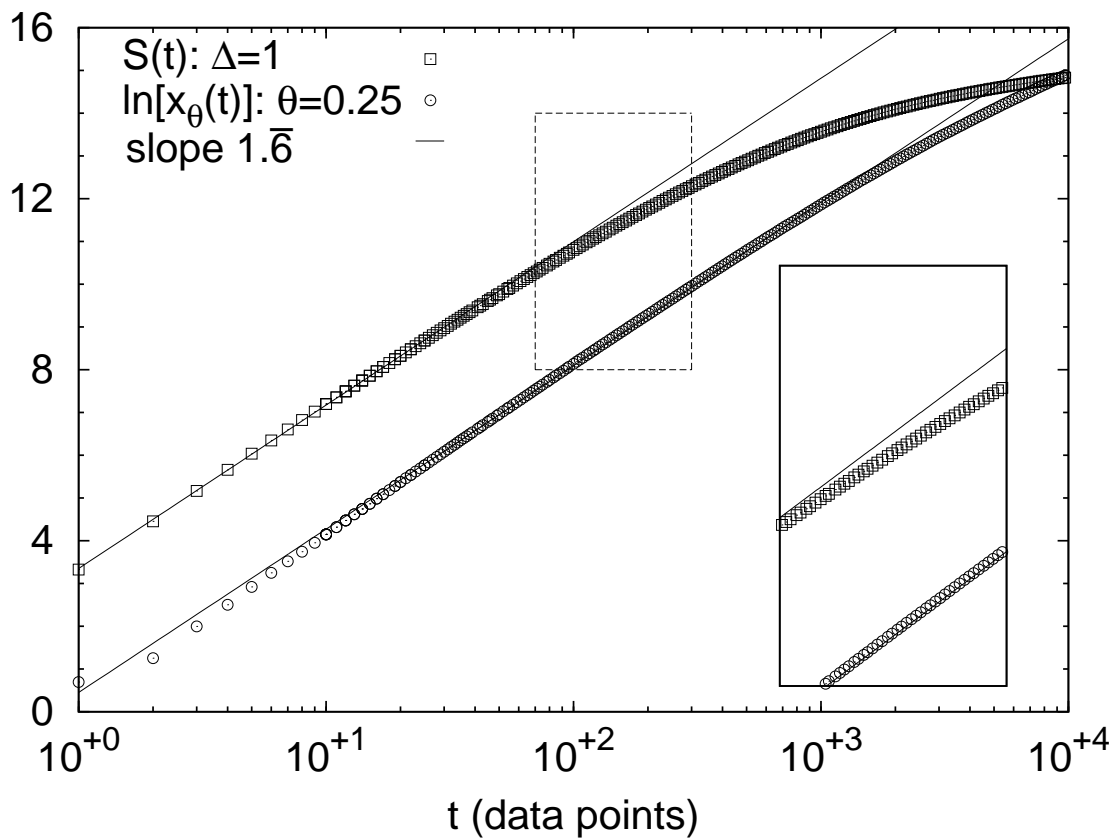}
\caption{Figure 1: Probability Flux Analysis (PFA) and the Diffusion Entropy
Analysis (DEA) are graphed for $10^{7}$ computer-generated data points using
(\protect\ref{plaw}) with a L\'{e}vy flight of scaling index $\protect\delta =1.\bar{%
6}(\protect\mu =1.6)$.}
\label{figure1}
\end{figure}

In Figure 2a the \textit{pdf} is depicted at times $t=50$ and $t=10^{3}$,
with the latter vertically displaced for visual clarity. In the vicinity of
the first time both the DEA and PFA calculations have the same slope as that
of theory, whereas in the vicinity of the latter time the DEA calculation
significantly deviates from the theoretical curve and the PFA calculation
does not. The \textit{pdf} at $t=50$ loses its crispness for large values of
the variate where the process becomes undersampled and therefore fluctuates
significantly from value to value. On the other hand, the \textit{pdf} at $%
t=10^{3}$ is undersampled throughout the domain of the distribution and
therefore the weight of successive values of the histogram are quite noisy.
This is what is meant by "running out of statistics" in the DEA calculation
in Figure 1; there is insufficient statistics to reduce the noise in the
histogram. The DEA calculation therefore is more useful at early times where
the histogram for the \textit{pdf} is more reliable. At latter times the
larger values of the variate become increasingly more important and the
histograms becomes increasingly less reliable. However, this lack of
reliability\ in the histograms has no apparent influence on the PFA since
slope determination with this latter method is done only using those values
of the variate below the percentage cut off $\theta$ and not on a faithful
reproduction of the \textit{pdf} in the entire range of the variate.

Another way of comparing the distribution at different times is to examine
the survival probabilities. L\'{e}vy distributions (with the exception of
the Gaussian case $\alpha =2)$ have inverse power-law tails with index $%
\alpha +1.$ Thus, the corresponding survival probability has an inverse
power-law tail with index $\alpha$. In Figure 2b, we see that at $t=50$ the
slope of the inverse power-law survival probability ($\alpha =\mu -1=0.6$ in
our case) coincides with the theoretical curve over multiple decades of
variate values. The extended inverse power-law region of the survival
probability is indicative of the quality of the \textit{pdf} depicted in
Figure 2a. On the other hand, the $t=10^{3}$ survival probability does not
have a region that coincides with the theoretical inverse power law for any
values of the variate. This lack of scaling is consistent with what is
observed from the DEA calculation in Figure 1. The empirical survival
probability is not sufficiently robust to detect the scaling in the data at
late times. However the survival probability is statistically robust at the $%
1-\theta $ level and thus the PFA method continues to detect the scaling.

\begin{figure}[ht]
\includegraphics[angle=-90,width=1.0\linewidth]{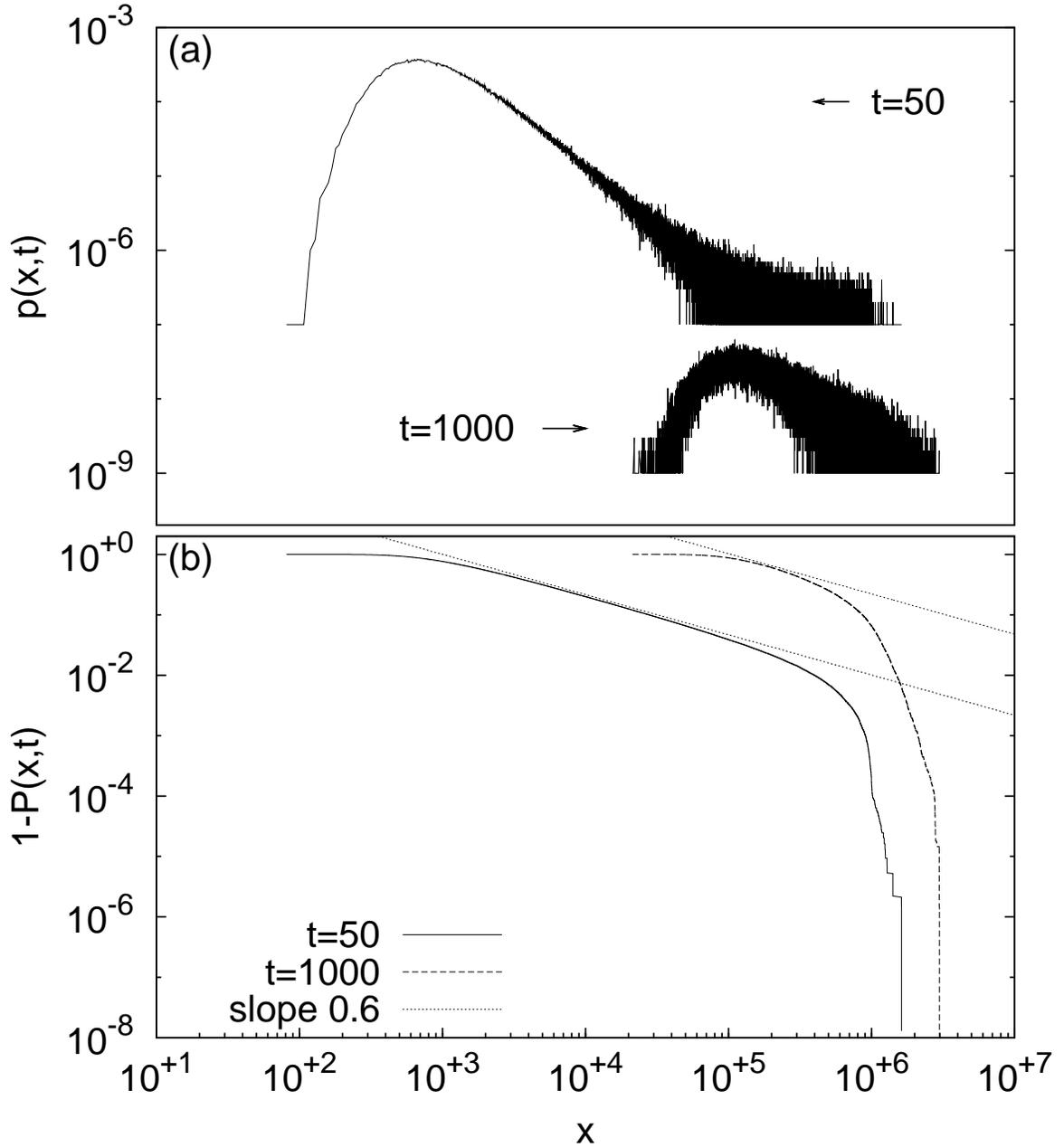}
\caption{Figure 2:(a) The \textit{pdf} is given at times $t=50$ and 10$^{3}.$%
It is evident that $p(x,t)$ is better at earlier times indicating that this
is where the DEA will be most useful. (b) The survival probability at the
two times is depicted. At $t=50$ the survival probability is an inverse
power law for more than two decades, whereas at $t=10^{3}$ there is
essentially no region where a theoretical inverse power law is detected.}
\label{figure2}
\end{figure}

To drive home this point it is useful to examine how the PFA depends on the
choice of $\theta$. In Figure 3 we plot the residue of the PFA analysis for
four widely spaced values of the fraction $\theta $ using the same
computer-generated data of Figure 1 and 2 ( a L\'{e}vy flight index $\alpha $%
$=$1.$\bar{6}$ ; $\mu $$=$1.6). The calculation tracking the theoretical
slope for the longest time $t=10^{4}$ has $\theta =0.05$. As the "time" t
increases the number of trajectories $N-t+1$ available in the STE decreases.
Therefore the larger $\theta $ the earlier in time the cumulative (survival)
probability $(1-\theta )$ becomes statistically unreliable. The divergence
of the first and second moments in the L\'{e}vy flight considered here makes
this loss of statistical robustness even more dramatic. A similar effect
will also occur if very small values of $\theta $, for example, $\theta
<<0.01$, are adopted. In fact, although the support of the \textit{pdf} is $%
[0,\infty ],$ the left border of the support of its numerical approximation
can be several orders of magnitude larger as shown in Figure 2

\begin{figure}[ht]
\includegraphics[angle=-90,width=1.0\linewidth]{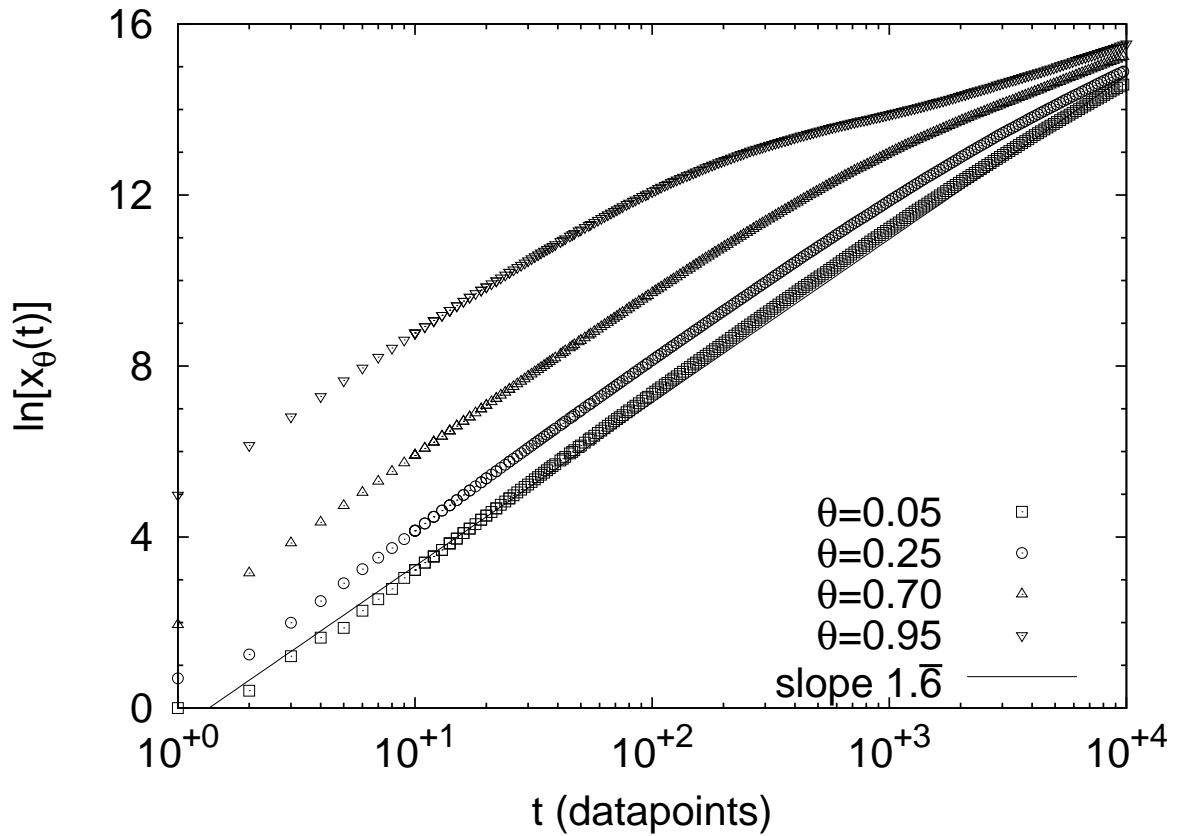}
\caption{Figure 3: The PFA calculation is graphed for $10^{7}$
computer-generated data points using (\protect\ref{plaw}) with a L\'{e}vy
flight of index $\protect\alpha =1.\bar{6}(\protect\mu =1.6)$ for a number
of different values of $\protect\theta$. }
\label{figure3}
\end{figure}

\subsection{L\'{e}vy walks}

A random walk is a stochastic process $X(t)$ where the distance covered by
the walker, the value of the variate, in a finite time is limited. The
restriction on the walk is a consequence of the physical relation between
taking a step of a given size and the time required to take such a step \cite%
{shlesinger}. Hence, at any given time $t$ the walk \textit{pdf} is bounded
and all the moments are finite. A L\'{e}vy walk has a \textit{pdf} between
the fastest walkers that is approximately equal, after a transient, to a L%
\'{e}vy distribution. L\'{e}vy walks can be generated in a number of
different ways, for example, using chaotic intermittent maps \cite%
{zumoklafter}, or, as we will do herein, using inverse power-law distributed
random numbers. We consider a type of L\'{e}vy walk called the Symmetric
Velocity Model (SVM) \cite{zumoklafter}. In the SVM walk, the velocity of
the random walker can only assume two values/states, here taken to be $\pm $%
1. The velocity of the walker remains in a given state for an interval of
time of random duration $\tau $ distributed according to the waiting-time
distribution density $\psi (\tau )$ given by the inverse power-law
distribution (\ref{plaw}) with index $\mu \in [2,3]$. After waiting
in a given state for a time $\tau$, that is, traveling at a constant
velocity for the specified time interval, a coin is tossed to determine the
new value of the velocity; and a new $\tau$ is extracted from the
distribution $\psi (\tau )$ to determine the duration of this new velocity
value. The pdf of the SVM walk at time $t$ is bounded between [$-t,t$] and
can be approximated ($t>>$ mean time of $\psi (t))$ by the scaling
expression Eq.(\ref{scaling}) with F(y) being a symmetric L\'{e}vy stable
distribution with shift parameter given by $\gamma =0,$ the skewness
parameter $\eta =0$, L\'{e}vy index $\alpha =\mu -1$, and by using the
scaling function $\beta (t)=t^{1/\alpha }$ \cite{dea,zumoklafter}.

For the numerical implementation of the SVM walk, we extract 10$^{6}$ random
waiting times $\tau $ according to the inverse power-law distribution with
index $\mu =2.5$ and location parameter $B=1$. Then we consider the
transformation $\tau \rightarrow \lbrack \tau ]+1$ where [.] is the integer
part of the term in brackets. This transformation creates a sequence of
integers \{$\tau _{k}>0\}$, which are inverse power-law distributed with the
same index $\mu $ and approximately the same location parameter B of the
original sequence. To assign the velocity states we generate a 10$^{6}$ long
sequence of random coin tosses: \{$v_{j}=\pm 1\}$. Finally we use the
couples $\{\tau _{j},v_{j}\}$ to create the sequence $\xi _{j}$ of
increments (the velocity of the walker) of the stochastic variable \textit{X}
and consequently the sequence $X_{j}$, which is processed using the DEA
algorithm.

In Figure 4 the PFA calculations for the four different values of $\theta $
are compared with the DEA calculation. After an initial transient ($t\sim
10^{2})$, the DEA calculation tracks the theoretical straight line ($%
S(t)\propto \delta \ln t$ with $\delta =1/(\mu -1)=0.6)$ up to a time $%
t\lesssim 10^{4}$ after which the DEA begins to run out of statistics. The 
\textit{pdf} of the SVM walk is symmetric, however the numerical evaluation
may not be symmetric, and at any time step we subtract the numerical mean
before calculating the value of $x_{\theta }(t)$, which satisfies Eq. (\ref%
{pfadefinition1}). Moreover due to the symmetry of the SVM walk $x_{\theta
=0.5}(t)=0,$ and $x_{\theta _{1}}(t)=-x_{\theta _{2}}(t)$ if $\left\vert
\theta _{1}-0.5\right\vert =\left\vert \theta _{2}-0.5\right\vert ,$ that
is, if the two different values of the parameter $\theta $ are symmetric
with respect to 0.5. Therefore, we limit ourselves to doing the PFA
calculations for $\theta >0.5$. The results depicted in Figure 4 indicate
that PFA tracks the therortical straight line for an additional decade
beyond that of DEA. The results are approximately independent of the
particular value of the cut off fraction $\theta $, although larger values
show less wiggly behavior than do smaller ones; an effect seemingly at odds
with those found in the previous case of the L\'{e}vy flight. The rationale
for this effect is the following. For the SVM L\'{e}vy walk, $x_{\theta
=0.5}(t)=0$ for all \textit{t}, however for the numerical calculation this
is not true and a plot of $x_{\theta =0.5}(t)$ reveals a fluctuating value.
The fluctuations increase in intensity as the time t increases since the
number of trajectories in the STE decreases as $N-t+1$, and the \textit{pdf}
becomes less symmetric. The cut off location $x_{\theta }(t)$ for values of $%
\theta $ closer to 0.5, such as 0.6 are more affected by this type of noise
than the cut off location for larger values of $\theta $ such as 0.9. The
overall effect is that although in theory $x_{\theta =0.6}(t)$ should be
more robust than $x_{\theta =0.9}(t),$ in practice, the latter is "crisper"
than the former. 
\begin{figure}[th]
\includegraphics[angle=-90,width=1.0\linewidth]{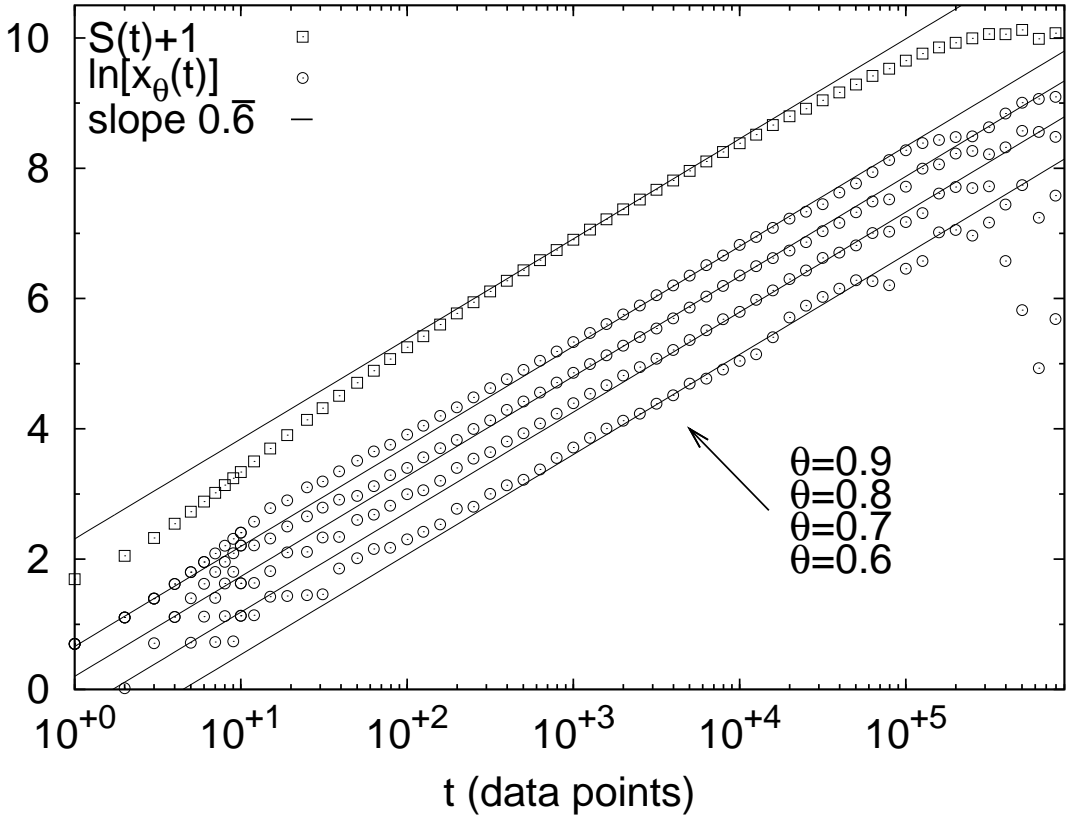}
\caption{Figure 4: The DEA and PFA calculations for a L\'{e}vy walk for four
values of the fraction of the data included in the calculation.}
\label{figure4}
\end{figure}

\subsection{EEG records}

In the previous two subsections we compared the results of using the DEA and
PFA methods to determine the known scaling properties of computer-generated
data sets having diverging and finite second moments, respectively. Now we
turn our attention to experimental data sets whose unknown scaling
properties we wish to determine. One such data set of both historic and
contemporary interest is that of the electroencephalogram (EEG) depicting
the erratic dynamics of the human brain. The observed scaling in EEG time
series is not as straightforward as observed in other less complex
phenomena. Various measures other than the standard deviation and spectrum
have been introduced into the study of EEG time series, each one stressing a
different physiologic property thought to be important in representing the
brain's dynamics. Most recently the DEA method has revealed a rather
remarkable behavior of the single channel EEG time series. Specifically the
failure of the EEG signal to scale: e.g.: Hwa et al.\cite{hwa02} applied DFA
to the series of increments of EEG increments and found a bi-scaling regime.
Ignaccolo \textit{et al}.~\cite{massiEEG} argues that the EEG signal during
resting activity can be modeled using a dissipative linear dynamic process $%
X(t)$, i.e., an Ornstein-Uhlenbeck process, with a quasi-periodic driver
having a random amplitude and frequency and an additive random force $\eta
\left( t\right) $ which is a delta correlated Gaussian process of strength $%
\sigma $. Latka \textit{et al}.~\cite{mireknow} shows how the model proposed
in \cite{massiEEG} explains the bi-scaling regime observed by Hwa et al~\cite%
{hwa02} and why this is not a \textquotedblleft real\textquotedblright\
algebraic scaling regime (satisfying Eqs.~(\ref{scaling}) and (\ref%
{algebraicscaling})) but is an artefact of DFA.

\begin{figure}[th]
\includegraphics[angle=-90,width=1.0\linewidth]{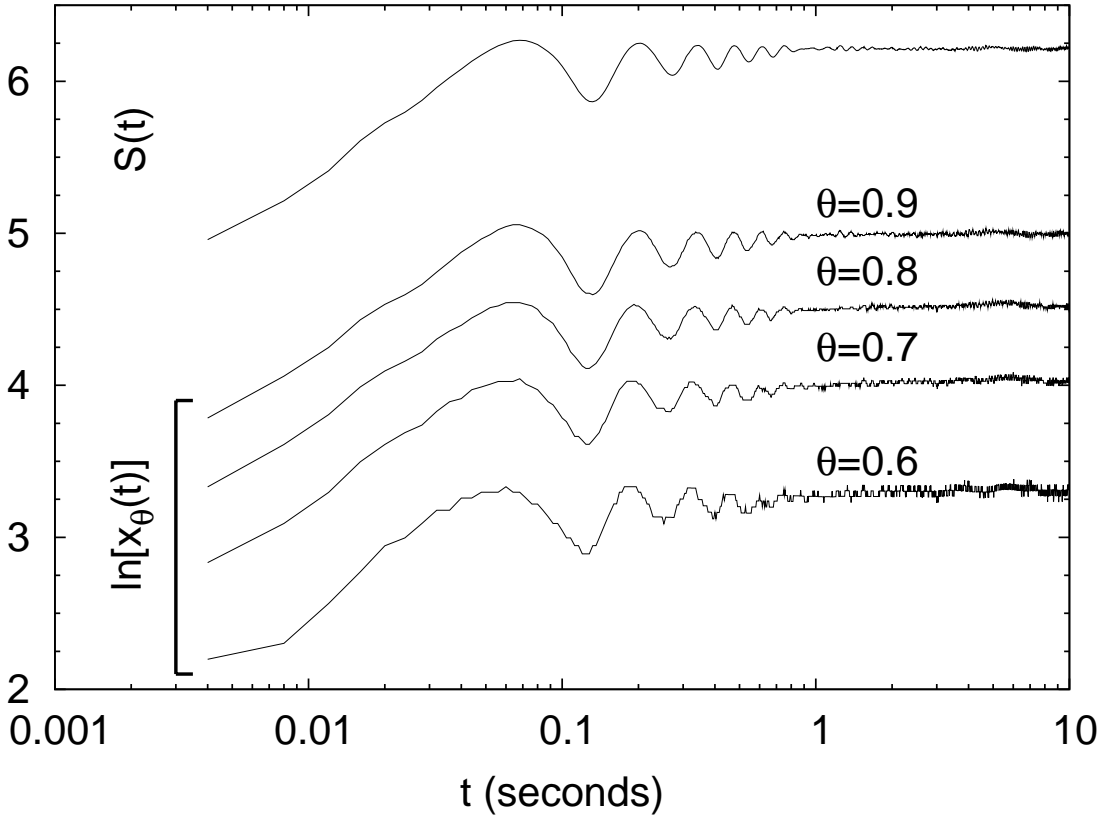}
\caption{The DEA processing of the EEG data from a single channel in the
occipital lobe is labeled $S(t)$. The equivalent PFA processing of the same
data using four different values $\protect\theta $ is also shown.}
\label{figure5}
\end{figure}

Fig.~\ref{figure5} shows the DEA for an EEG channel under the closed eyes
resting condition. The time $t$ is expressed in seconds with 1 second
corresponding to 250 data samples. The observed saturation is the results of
dissipative linear dynamics while the decaying oscillations are produced by
the random periodic forcing: the alpha rhythm which is the well known wave
pattern present in EEG under the closed eyes resting condition. Different
trajectories of the STE (\ref{pseudogibbs}) experience wave packets of
different amplitude and frequency. The typical duration of a wave packet is $%
\sim $0.5s \cite{massiEEG}. The mixture of wave packets results in a pattern
of destructive interference. The larger the time $t$, the wider the spectrum
of different amplitudes and frequencies present in the trajectory $X_{k}(t)$
of Eq.~(\ref{pseudogibbs}), and the more intense is the interference. This
mechanism explains the observed decaying oscillation for the information
entropy $S(t)$. The apparent period $\sim $0.013s is just the amplitude
weighted average of $\alpha $$-$wave packet periods occurring in the
particular EEG record examined. Also plotted in Fig.\ref{figure5} is the PFA
for different values of the cut off fraction $\theta $. We see how PFA
reproduces all the characteristic observed for DEA. Note that EEG records
are almost symmetric so that the PFA analysis for $\theta $$<$0.5 is just
the mirror image of the one shown, as is the L\'{e}vy SVM walk of the
previous section, of the one relative to $\theta $$>$0.5. This also explain
why the results are increasingly \textquotedblleft crisp\textquotedblright\
when moving from $\theta $$=$0.6 to $\theta $$=$0.9 in Fig. 5. 

\section{Conclusion}

\label{conclusion} The PFA calculates the \textquotedblleft cut
off\textquotedblright\ location $x_{\theta }$ which encompasses a fraction $%
\theta $ of the \textit{pdf} of the stochastic process $X(t)$. As the 
\textit{pdf} evolves in time so does $x_{\theta }$ in order to always
encompass the same fraction of the distribution. In this sense, PFA is a
\textquotedblleft volume\textquotedblright\ preserving transformation. The
volume preserving transformation for a scaling distributions is, aside from
a constant multiplicative factor, the scaling function itself (\ref%
{pfascaling2}). Therefore PFA can be a scaling detection tool. The power of
this method is that the volume preserving transformation is statistically
more robust than any method based on a detailed knowledge of the \textit{pdf}
such as DEA. The PFA with cut off fraction $\theta $ is as statistically
robust as the cumulative distribution $\mathcal{P}(x,t)$ at level $\theta $
or the survial distribution $(1-\mathcal{P}(x,t))$ at level $(1-\theta )$.
For a bounded signal with a \textquotedblleft well-behaved
distribution\textquotedblright\ (no inverse power law with infinite first or
second moment) such as the EEG data the statistical advantage of the PFA
over DEA may not be so evident (Fig.~\ref{figure5}). However, as soon as we
depart from the realm of well-behaved distributions the statistical
advantage of PFA becomes apparent. For the SVM L\'{e}vy walk the agreement
with the theoretical curve is extended by one decade (Fig.~\ref{figure4}).
SVM L\'{e}vy walks have bounded \textit{pdf's} but a non well-behaved
distribution is hidden in this stochastic process: the \textit{pdf} of the
waiting times (\ref{plaw}) with index $\mu $$\in $]2,3[, which has an
infinite variance, and is the reason why a L\'{e}vy distribution appears in
the region between the bounding sites. The statistical advantage of PFA over
DEA is even more evident in the case of L\'{e}vy flights where the \textit{%
pdf} is unbounded with infinite first and second moments. In this latter
case, almost two decades are gained using one method rather than the other
(Fig.~\ref{figure1} and \ref{figure3}).

Aside from scaling detection, the results of Fig.~\ref{figure5} relative to
the EEG record, where there is no algebraic scaling, shows that PFA can be
used to investigate the dynamics of a stochastic process $X(t)$. In fact,
the details of the dynamics generating the signal $X(t)$ are somewhat
reflected in the volume preserving transformation performed by PFA. From
these results the inescapable conclusion is that the PFA method \ is
superior to the DEA and ought to replace it. Moreover, since we have shown
elsewhere \cite{massiEEG} that DEA is preferable to detrended fluctuation
analysis (DFA), we must further conclude that PFA replace DFA as well.

M. I. and B. J. W. thank the Army Research Office for support of this research.


\begin{thebibliography}{99}
\bibitem{dereport} In most pratical application there is no difference
between the overlapping and non-overlapping procedures aside from a better
statistical result for the overlapping one.
\end{thebibliography}
\end{document}